\documentclass[12pt,preprint]{aastex62}

\usepackage{amsmath}

\newcommand{\IRIS}{{\em IRIS}}

\newcommand{\GOES}{{\em GOES}}

\newcommand{\siiv}{Si {\sc iv}}
\newcommand{\fexxi}{Fe {\sc xxi}}

\shorttitle{}
\shortauthors{}

\begin{document} 

\title{\IRIS~\siiv~Line Profiles at Flare Ribbons as Indications of Chromospheric Condensation}

\author{Ke Yu}           
\affiliation{School of Astronomy and Space Science, Nanjing University, Nanjing 210023, China}
\affiliation{Key Laboratory for Modern Astronomy and Astrophysics (Nanjing University), Ministry of Education, Nanjing 210023, China}
\author{Y. Li}
\affiliation{Key Laboratory of Dark Matter and Space Astronomy, Purple Mountain Observatory, Chinese Academy of Sciences, Nanjing 210033, China}
\author{M. D. Ding}           
\affiliation{School of Astronomy and Space Science, Nanjing University, Nanjing 210023, China}
\affiliation{Key Laboratory for Modern Astronomy and Astrophysics (Nanjing University), Ministry of Education, Nanjing 210023, China}
\author{D. Li}
\affiliation{Key Laboratory of Dark Matter and Space Astronomy, Purple Mountain Observatory, Chinese Academy of Sciences, Nanjing 210033, China}
\author{Yi-An Zhou}           
\affiliation{School of Astronomy and Space Science, Nanjing University, Nanjing 210023, China}
\affiliation{Key Laboratory for Modern Astronomy and Astrophysics (Nanjing University), Ministry of Education, Nanjing 210023, China}
\author{ Jie Hong}           
\affiliation{School of Astronomy and Space Science, Nanjing University, Nanjing 210023, China}
\affiliation{Key Laboratory for Modern Astronomy and Astrophysics (Nanjing University), Ministry of Education, Nanjing 210023, China}
\email{dmd@nju.edu.cn}

\begin{abstract}
We present temporal variations of the \siiv~line profiles at the flare ribbons in three solar flares observed by the {\em Interface Region Imaging Spectrograph} (\IRIS). In the M1.1 flare on 2014 September 6 and the X1.6 flare on 2014 September 10, the \siiv~line profiles evolve from wholly redshifted to red-wing enhanced with the flare development. However, in the B1.8 flare on 2016 December 2, the \siiv~line profiles are wholly redshifted throughout the flare evolution. We fit the wholly redshifted line profiles with a single Gaussian function but the red-asymmetric ones with a double Gaussian function to deduce the corresponding Doppler velocities. In addition, we find that hard X-ray emission above 25 keV shows up in the two large flares, implying a nonthermal electron beam heating. In the microflare, there only appears weak hard X-ray emission up to 12 keV, indicative of a thermal heating mostly. We interpret the redshifts or red asymmetries of the \siiv~line at the ribbons in the three flares as spectral manifestations of chromospheric condensation. We propose that whether the line appears to be wholly redshifted or red-asymmetric depends on the heating mechanisms and also on the propagation of the condensation.
\end{abstract}


\section{Introduction}

Solar flares are events of rapid energy release in the solar atmosphere that are observed as sudden brightenings particularly in chromospheric and transition region (TR) spectral lines. It is suggested that the flare energy is initially released in the corona, which heats the local plasma and accelerates charged particles. Then the energy is transported downward through either thermal conduction or non-thermal particle beams, and is mostly deposited in the chromosphere. The chromosphere is heated to generate an enhanced radiation that is seen as flare ribbons. During the heating process, hard X-ray (HXR) emission can be produced by non-thermal electrons impacting the ambient atoms. In particular, heating of the chromosphere results in an excessive pressure, which can drive the plasma flowing upward, a phenomenon known as chromospheric evaporation \citep{neup68, hira74, acto82}. There have been suggested two types of chromospheric evaporation, gentle and explosive ones \citep{fish85b, mill06a, mill06b}. For the explosive evaporation, the upward momentum is large enough that at the same time induces a noticeable downward compression front, known as chromospheric condensation \citep{fish87, canf90}. In the gentle evaporation, however, obvious signatures of condensation have not been observed.\\

Chromospheric evaporation and condensation are mostly manifested in the Doppler shifts of spectral lines of relatively high temperatures and those of relatively low temperatures, respectively. Extensive efforts have been devoted to the spectral observations and spectral diagnostics on the chromospheric evaporation and condensation. Generally speaking, the most prominent spectral feature of the hot emission lines, such as Ca {\sc xix}, Fe {\sc xix}, Fe {\sc xxi}, Fe {\sc xxiii}, and Fe {\sc xxiv} lines, is a blueshift or blueshifted component with a Doppler velocity of tens to hundreds of km s$^{-1}$, which corresponds to chromospheric evaporation \citep[e.g.,][]{anto82, bros04, bros09b, bros09a, mill09, raft09, wata10, trip12, dosc13, youn15}. For the relatively cool lines formed in the chromosphere and TR, such as H$\alpha$, Ca {\sc ii}, He {\sc i}, and O {\sc v} lines, they usually exhibit a redshift or red asymmetry with a Doppler velocity of a few tens of km s$^{-1}$, which is caused by chromospheric condensation \citep[e.g.,][]{dela92, falc92, neid93, ganw93, wuls94, ding95, ding96, kuri17}. In fact, simultaneous appearance of both blueshifts and redshifts in lines of different temperatures, indicative of an explosive evaporation, has been observed at flare ribbons by different instruments. Based on the data from the Coronal Diagnostic Spectrometer \citep[CDS;][]{harr95} on board the {\em Solar and Heliospheric Observatory}, \citet{teri06} reported strong blueshifts in the Fe {\sc xix} line and redshifts in the O {\sc v} and He {\sc i} lines. With a higher spatial resolution and cadence as well as a wider temperature coverage, the Extreme-ultraviolet Imaging Spectrometer \citep[EIS;][]{culh07} on board {\em Hinode} \citep{kosu07} also detected blueshifts and redshifts simultaneously in different lines that can be explained in terms of explosive evaporation \citep{wata10, bros16, liyd11, dosc13}. In particular, it has been found that some emission lines formed at relatively high temperatures also show redshifts \citep[e.g.,][]{mill09, chen10, youn13}. \\

In addition, the {\em Interface Region Imaging Spectrograph} \citep[{\em IRIS};][]{depo14} has provided plenty of spectral data with a high spatial and temporal resolution since its launch in 2013. It offers a good opportunity to study chromospheric evaporation and condensation in more details. For example, chromospheric evaporation has been detected as a whole blueshift in the hot Fe {\sc xxi} 1354.08 {\AA} line \citep[e.g.,][]{poli15, poli16, sady15, dudi16}. On the other hand, chromospheric condensation has been observed in a few important cool lines. \citet{tian14} found that the cool O {\sc iv}, \siiv, C {\sc ii} and Mg {\sc ii} lines display evident redshifts at the flare loop footpoints. \citet{bros15} studied the \siiv~1402.77 {\AA} line and reported downward velocities that are consistent with chromospheric condensation. \citet{grah15} revealed sudden and strong condensation downflows from the red asymmetry of the Mg {\sc ii} 2791.59 {\AA} line. Moreover, through a comparison between the blueshifted Fe {\sc xxi} 1354.08 {\AA} line and redshifted C {\sc i} 1354.29 {\AA} line, \citet{lidn15} studied explosive evaporation in two X1.6 flares, which is mainly driven by electron beam heating. By checking the temporal evolution of the line features, \citet{sady16} found a delay of the maximum blueshift of the Fe {\sc xxi} 1354.08 {\AA} line relative to the maximum redshift of the C {\sc ii} 1334.53 {\AA} line. \\

 The two resonance \siiv~lines at 1393.75 and 1402.77 {\AA} observed routinely by {\em IRIS} have been used in a number of studies on chromospheric condensation \citep[reviewed by][]{liyd19}. \citet{tian14} reported Si {\sc iv} 1402.77 {\AA} redshifts with a velocity of about 50 km s$^{-1}$ at the loop footpoint produced by chromospheric condensation. \citet{liyd15} displayed red-asymmetric \siiv~1402.77 {\AA} line profiles at the flare ribbons caused by condensation plasma. \citet{tian15} also reported asymmetric \siiv~1402.77 {\AA} line profiles with an enhanced red wing but not entirely redshifted, which indicate a larger speed (about 100 km s$^{-1}$) of chromospheric condensation than previously thought. In addition, \citet{bran15} showed the \siiv~line profiles with a red asymmetry or even two separate peaks within the ribbon of an M-class flare. \citet{warr16} found that the \siiv~1402.77 {\AA} line exhibits a stronger redshifted component but a weaker stationary component. Considering the line profile shape, \citet{leek17} used a multiple-component Gaussian function to fit the \siiv~1402.77 {\AA} line and obtained the intensity, Doppler velocity and line width for each component. Furthermore, \citet{liyk17} studied an X-shaped flare and found that the \siiv~1402.77 {\AA} line changes significantly in shape at different locations. They considered that the \siiv~line is either wholly shifted or asymmetric, in relation to different kinds of energy deposition. \citet{zhan16} discussed periodic chromospheric condensation through analyzing the \siiv~1402.77 {\AA} line at different locations in a C3.1 circular ribbon flare. \citet{tian18} discussed repeated chromospheric condensation and energy injection in the form of nonthermal electrons by analyzing the redshifts of the \siiv~1402.77 {\AA}, Mg {\sc ii} 2791.59 {\AA}, and Mg {\sc ii} k 2796.35 {\AA} lines. \\

In this paper, we focus on the {\em IRIS} \siiv~1402.77 {\AA} line and study its temporal variations in three flares. In particular, we perform different methods to the line profiles with different shapes to obtain the physical parameters. By comparing the line features with the HXR emission, we find that the whole redshift and red asymmetry of the \siiv~line may correspond to different flare heating modes. In Section \ref{sec-instr}, we describe the instruments and data reduction. Then we describe the methods of moment analysis and Gaussian fitting in Section \ref{thi-met}. Section \ref{fou-res} presents the results for the three flares, and Sections \ref{fiv-dis} and \ref{six-sum} provide discussions and conclusions, respectively.\\

\section{Instruments and Data Reduction} 
\label{sec-instr}

The data used in this study come from a few instruments that are described in the following. We select three flares (listed in Table \ref{tab-table}) for study that possess the following observational characteristics: a variety of flare magnitude (both large and small flares), high enough time cadence (say, $<$10 s), full spatial coverage (flare ribbons) and temporal coverage (rise phase of the flare), and HXR observations.  \\

{\em IRIS} provides high-resolution slit-jaw images (SJIs) as well as spectral data via a slit. The slit has a width of $0.^{''}33$ and a maximum length of 175$^{''}$. For SJIs, the maximum field of view  is $175^{''}\times175^{''}$ and the pixel size is $0.^{''}166$. There are two observation modes to acquire the slit spectra, either raster scan or sit-and-stare, for the latter of which the time cadence can be as high as 1--2 s. Note that all the three flares under study are observed with a sit-and-stare mode. The relevant observational parameters including the time cadence and the pixel size along the slit are given in Table \ref{tab-table}. The level 2 data ready for scientific use are analyzed here. \\


\begin{deluxetable*}{lcccccccccl}[b!]
\tablecaption{List of the flares analyzed in the paper 
\label{tab-table}}
\tablenum{1}
\tablehead{
\colhead{Event} & \colhead{Date of} & \colhead{\GOES} & \colhead{Start} & \colhead{Peak} & \colhead{End} & \colhead{Cadence\tablenotemark{1}} & \colhead{Pixel size\tablenotemark{2}} & \colhead{HXR emission\tablenotemark{3}}  \\
\colhead{Number} & \colhead{Observation} & \colhead{Class} & \colhead{Time} & \colhead{Time} & \colhead{Time} & 
\colhead{(s)} & \colhead{($^{''}$)} & \colhead{(keV)}
}
\startdata
Flare 1      &  2014-09-06  &  M1.1            &  16:50  &  17:09  &  17:22   &  9.5  &  0.166  & {\em RHESSI} 25--50  \\
Flare 2      &  2014-09-10  &  X1.6             &  17:21  &  17:45  &  17:45   &  9.5  &  0.166  & {\em Fermi} 51--102 \\
Flare 3      &  2016-12-02  &  B1.8\tablenotemark{4}            &  16:14  &  16:24  &  16:36   &  1.7  &  0.333   & {\em RHESSI} 6--12  \\
\enddata
\tablenotetext{1}{ This is the time cadence of the {\em IRIS} spectra that were obtained with a sit-and-stare mode.}
\tablenotetext{2}{ The pixel size is for {\em IRIS} SJIs. Note that there is a spatial binning for flare 3.}
\tablenotetext{3}{ It represents the highest energy band from {\em RHESSI} or {\em Fermi}, at which evident HXR emission was detected.}
\tablenotetext{4}{ There is no record of the flare in the {\em GOES} list. The flare class is determined based on the peak of 1--8 {\AA} flux.}
\end{deluxetable*}
 
\IRIS\ spectral windows include lines formed over a wide temperature range from about 4500 K to 10 MK, thus reflecting physical properties from the lower atmosphere to the corona. We focus on the TR \siiv~1402.77 {\AA} line \footnote{This line may suffer from a blending with some relatively weak lines \citep[e.g.,][]{doyl92}. However, the effects of the blending on the \siiv~line intensity as well as Doppler velocity are trivial and can be safely neglected based on theoretical calculations.} that has a formation temperature of about 10$^{4.9}$ K. Generally, the \siiv~line can be considered as optically thin; thus we can easily derive the Doppler velocity from the line profiles. In order to determine the reference wavelength of the \siiv~line, we first apply the single Gaussian fitting to the line profiles in a quiet region before the flare and then make an average of the line centers. After the reference line center is determined like that, the Doppler shifts calculated from the lines in the flare region are purely flare-induced. The reference line centers determined here vary marginally for different events. For example, the reference wavelength is about 1402.7852 {\AA} for flare 1 and 1402.7910 {\AA} for flare 2. Note that in flare 3 (a microflare), the \siiv\ line  is too weak outside the flare region; thus we just use the theoretical value of 1402.7700 {\AA} as the reference wavelength. The uncertainty in Doppler velocity is estimated to be $\sim$5 km s$^{-1}$. \\

For HXR observations, flares 1 and 3 were captured by the {\em Reuven Ramaty High Energy Solar Spectroscopic Imager} \citep[{\em RHESSI};][]{linr02} and flare 2 by {\em Fermi} Gamma-ray Burst Monitor \citep[GBM;][]{meeg09}. {\em RHESSI} can image and observe the Sun in X-ray and $\gamma$-ray bands (3 keV--17 MeV) with an energy resolution of about 1 keV to 5 keV. The spatial resolution can be as high as $2.^{''}3$ and the temporal resolution is about 2 s or better. For flare 1, we reconstruct the 25--50 keV image using detectors 2--8 from 16:55:00 UT to 16:55:32 UT. For flare 3, however, the emission above 12 keV is very low, and thus we do not reconstruct the image here. {\em Fermi} GBM has an energy band ranging from 8 keV to 40 MeV. It has 12 detectors, whose viewing angles towards the Sun vary with time during their operation. During the period of flare 2, the viewing angles of detectors n2 and n4 towards the Sun are almost unchanged (about $60^{\circ}$). Here we choose the data from detector n2 for analysis. Note that in this work, we only plot the HXR light curve at the highest energy band that has evident emission for each of the flares (see Table \ref{tab-table}).  \\

We also use the data from some other instruments including the Atmospheric Imaging Assembly \citep[AIA;][]{leme12} on board the {\em Solar Dynamics Observatory} \citep[{\em SDO};][]{pesn12} and also the {\em Geostationary Operational Environmental Satellites} ({\em GOES}). AIA is designed to make full-disk imaging observations from multiple bands, which include seven EUV (94 {\AA}, 131 {\AA}, 171 {\AA}, 193 {\AA}, 211 {\AA}, 304 {\AA}, and 335 {\AA}), two UV (1600 {\AA} and 1700 {\AA}) and one white-light (4500 {\AA}) channels. The data from AIA are of high spatial resolution ($\sim$$1.^{''}2$) and high cadence (12 or 24 s). {\em GOES} has two soft X-ray bands, 0.5--4 {\AA} and 1--8 {\AA}. The 1--8 {\AA} flux is usually used to monitor the response of a flare. \\

\section{Methods}
\label{thi-met}

The \siiv~line formed in the TR is usually treated as optically thin, although it may have some opacity effects especially during flares \citep[e.g.,][]{math99, kerr19}. Considering that this line does not show any reversal or absorption feature in the line core and that the line profiles can be well fitted by a Gaussian function (with either a single or double components) in the three flares generally, we assume this line to be optically thin here. If there are no obvious dynamics in the atmosphere, this line can be well represented by a single Gaussian shape. However, this is not always the case in flares, especially when multi-velocity flows exist along the line of sight. Therefore, we need to perform different methods in different cases when quantitatively deriving the parameters from the line profiles, which include the moment analysis, a single-Gaussian fitting, and a multiple-Gaussian fitting. \\

\subsection{Moment Analysis}

The zeroth, first, and second moments, denoted by $I$, $\lambda_c$, and $\sigma$, respectively, are defined as following:

\begin{equation}
I=\int(f(\lambda)-f_0)\,d\lambda,
\end{equation}

\begin{equation}
\lambda_c=\left[ \int\lambda(f(\lambda)-f_0)\,d\lambda \right]/I,
\end{equation}

and

\begin{equation}
\sigma^2=\left[ \int(\lambda-\lambda_c)^2(f(\lambda)-f_0)\,d\lambda \right]/I,
\end{equation}
where $f(\lambda)$ is the observed line intensity at wavelength $\lambda$ and $f_{0}$ is the background intensity. \\

The zeroth moment $I$ corresponds to the integrated intensity over wavelength. The first moment $\lambda_c$ yields the centroid position of the line profile, from which the Doppler shift velocity is calculated by
\begin{equation}
v=c(\lambda_c-\lambda_r)/\lambda_r,
\end{equation}
where $c$ is the speed of light and $\lambda_r$ is the reference wavelength of the line. The second moment $\sigma$ is a measure of the line width reflecting the velocity dispersion of the plasma at the line formation layer. The moment analysis is robust and can be performed to all kinds of line profiles. However, it can only yield parameters on average (like the line shift and line width) that should be used with caution in particular when dealing with very asymmetric line profiles. \\

\subsection{Gaussian Fitting}

If the line profile shows a Gaussian shape, it can be fitted by the following function: 
\begin{equation}
f(\lambda)=I_p\exp \left[ -\frac{(\lambda-\lambda_c)^2}{(\Delta\lambda_D)^2} \right]+f_0,
\end{equation}
where $I_p$ is the peak intensity above the background and $\lambda_c$ is the central wavelength of the observed line profile. The parameter $\Delta\lambda_D$ refers to the Doppler width of the line profile.  \\

For some asymmetric line profiles, we also employ a double-Gaussian fitting. This assumes that the profile consists of two Gaussian-shaped components and a constant background \citep{hong16}, so that
\begin{equation}
f(\lambda)=I_1\exp\left[ -\frac{(\lambda-\lambda_1)^2}{(\Delta\lambda_1)^2} \right]+I_2\exp\left[ -\frac{(\lambda-\lambda_2)^2}{(\Delta\lambda_2)^2} \right]+f_0,
\end{equation}
where $I_1$ and $I_2$ are the peak intensities, and $\Delta\lambda_1$ and $\Delta\lambda_2$ are the Doppler widths of the two components, respectively. The parameters $\lambda_1$ and $\lambda_2$ refer to the central wavelengths of the two components, which can be used to calculate the Doppler velocity via Equation (4). \\ 

\begin{figure*}
\centering
\includegraphics[width=15cm, height=8 cm]{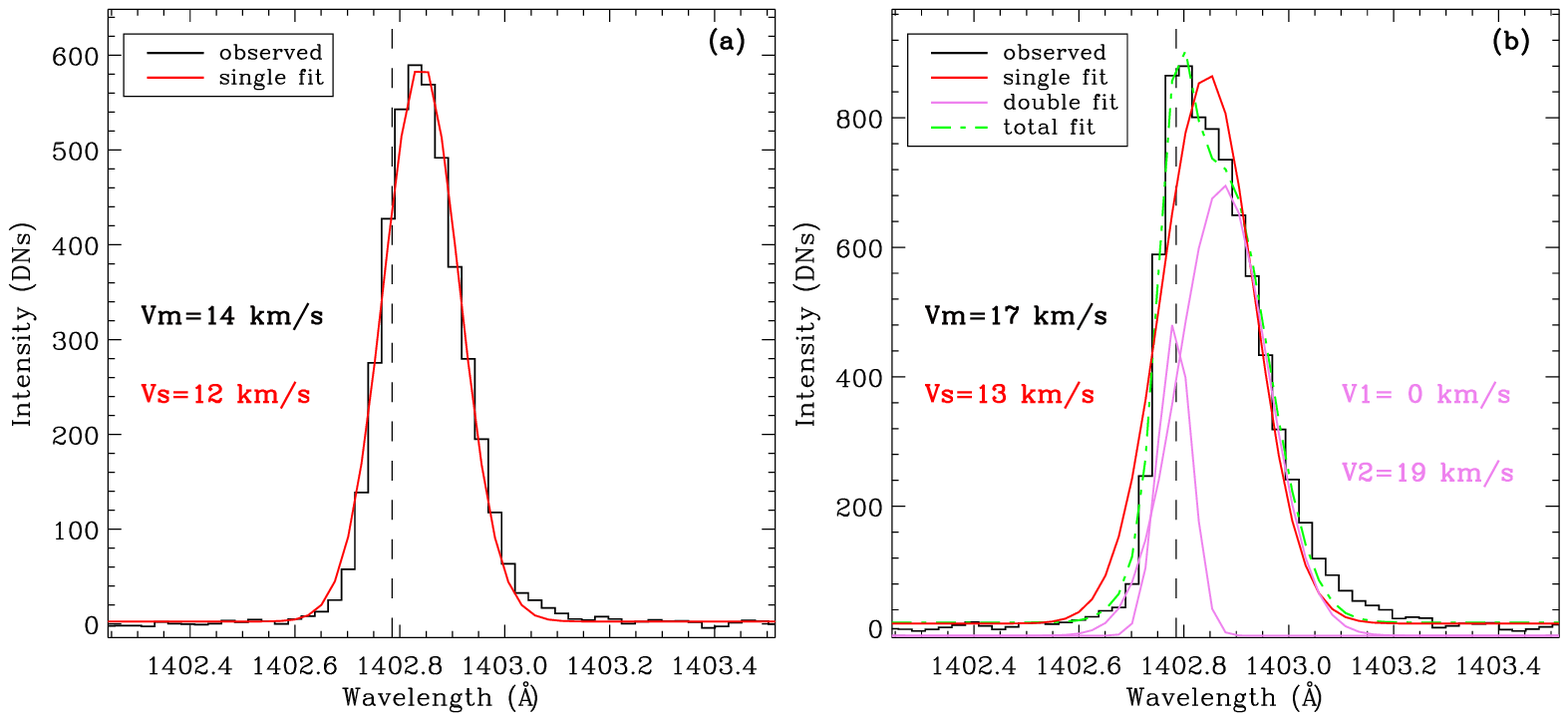}
\caption{{\small Two typical line profiles of \siiv~1402.77 {\AA}. The left panel is for a wholly redshifted profile (black), overplotted on which is the single Gaussian fitting (red). The right panel is for a red-asymmetric profile (black), overplotted on which are the single Gaussian fitting (red) and double-Gaussian fitting (green), as well as the two components (violet) for the latter. The vertical dashed line indicates the reference wavelength. The Doppler velocities are given in each panel: $v_{m}$ from the moment analysis, $v_{s}$ from the single Gaussian fitting, and $v_{1}$ and $v_{2}$ for the two components of the double-Gaussian fitting, respectively.  }}
\label{fig-met}
\end{figure*}

For the three flare events under study, we first make a moment analysis on the observed \siiv~line profiles to get the average parameters, and then adopt a single or double Gaussian fitting to further extract the line parameters more accurately. Two typical examples of the observed line profiles and their fitting curves are shown in Figure \ref{fig-met}. In the first case (Figure \ref{fig-met}(a)), the observed \siiv~line profile presents a good single Gaussian shape. Accordingly, we can see that the Doppler velocities derived from the moment analysis and the single Gaussian fitting are almost the same. Note that there still exists a tiny difference between the two velocities, which is supposed to be caused by a deviation from a purely symmetric line profile. In the second case (Figure \ref{fig-met}(b)), the observed line profile displays an evident enhancement at the red wing. It is obvious that the single Gaussian fitting is not appropriate for such a line profile with an evident asymmetry. Hence we adopt a double Gaussian function to fit the line profile, which yields a better and more precise result. It is seen that the first Gaussian component is relatively stationary, while the second one is redshifted, which contributes to the red-wing enhancement. The Doppler velocity of the second component is of course larger than the velocities derived from the moment analysis and single Gaussian fitting. This suggests that adopting a single Gaussian fitting may underestimate the redshift velocity of the \siiv~line profiles that show a significant red asymmetry. Quantitatively, we perform a double Gaussian fitting to those line profiles if the velocities derived from a single Gaussian fitting and the moment analysis differ by larger than $\sim$3 km s$^{-1}$. Note that there are very few line profiles that could be better fitted by a three-Gaussian function. However, considering that the third Gaussian component is quite weak in intensity, we still perform a double Gaussian fitting to those line profiles.

\section{Observations and Results}
\label{fou-res}

\subsection{Flare 1: the M1.1 flare on 2014 September 6}
\label{fou-res1}
\subsubsection{Observation Overview}
\begin{figure*}
\centering
\includegraphics[width=16cm, height=12cm]{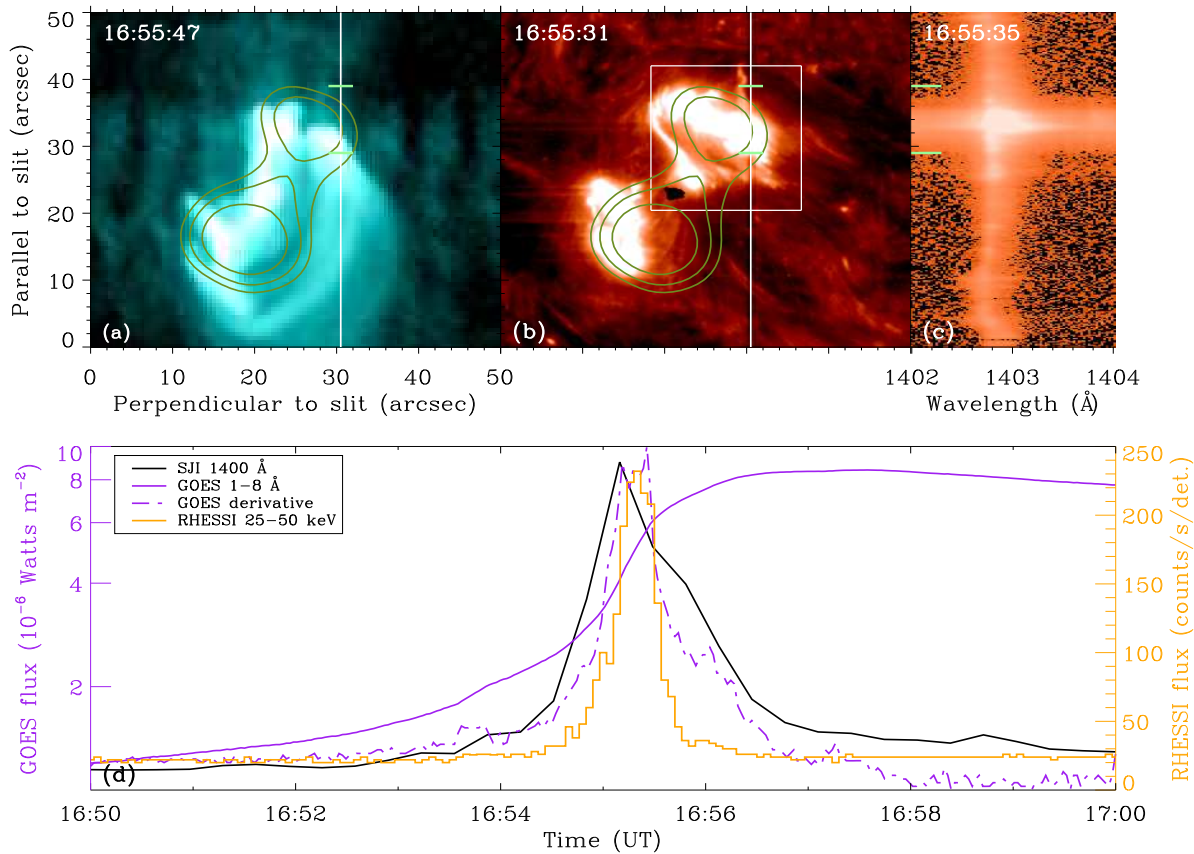}
\caption{{\small (a) {\em SDO}/AIA 131 {\AA} image taken at 16:55:47 UT on 2014 September 6 showing the morphology of flare 1. (b) {\em IRIS}/SJI 1330 {\AA} image taken at 16:55:31 UT. (c) {\em IRIS} slit spectra of the \siiv~1402.77 {\AA} line taken at 16:55:35 UT. Also plotted on panels (a) and (b) are the slit position (vertical white line) and the contours (olive) of {\em RHESSI} 25--50 keV HXR emissions (with contour levels of 8\%, 10\%, and 13\% of the peak emission). The two green bars enclose the part of the slit that is shown in Figure \ref{fig-mo1}. (d) Light curves of the SJI 1400 {\AA} intensity (black, arbitrary scale) averaged over the flare region (marked by the white square), {\em GOES} 1--8 {\AA} flux (purple solid) and its time derivative (purple dash-dotted), and {\em RHESSI} 25--50 keV flux (orange). }}
\label{fig-rhe}
\end{figure*}

The M1.1 flare on 2014 September 6 occurred in NOAA AR 12157. The flare started at 16:50 UT and ended at 17:22 UT. There exist two peaks in the {\em GOES} SXR light curve (about 16:56 UT and 17:09 UT), which imply two episodes of energy release in this flare \citep{tian15}. Here we mainly focus on the first SXR peak during the time period of 16:45--17:01 UT. This flare was well observed by {\em IRIS}, {\em SDO}/AIA, and {\em RHESSI}. The {\em IRIS} SJIs were taken at 1330 and 1400 {\AA} with a cadence of $\sim$19 s. From $\sim$16:55 UT, the exposure time of the SJIs changed from 8 s to 2.4 s. Figures \ref{fig-rhe}(a)--(c) show the AIA 131 {\AA} image, the SJI at 1330 {\AA}, and the spectra of \siiv~during the flare, with the {\em RHESSI} HXR sources at 25--50 keV overplotted on the former two panels. Note that the {\em RHESSI} HXR and AIA 131 {\AA} images have been rotated counterclockwise by 45$^{\circ}$ to fit the orientation of the {\em IRIS} SJIs. One can clearly see some flare loops in the AIA 131 \AA\ image as well as their footpoints (or flare ribbon) in the SJI at 1330 \AA\ that well match the HXR sources. The {\em IRIS} slit crossed some of the flare ribbons where prominent emission shows up in the \siiv\ spectra. Figure \ref{fig-rhe}(d) shows the {\em GOES} SXR, {\em RHESSI} HXR, and SJI 1400 {\AA} light curves. For comparison, we also plot the time derivative of the SXR flux. It is seen that the HXR emission and the SXR time derivative show a similar temporal evolution with peaks at $\sim$16:55 UT, implying the validity of the Neupert effect for this flare. The SJI 1400 \AA\ emission also peaks at nearly the same time within the time resolution of the observations. 

\subsubsection{Results}

\begin{figure*}
\centering
\includegraphics[width=16cm, height=12cm]{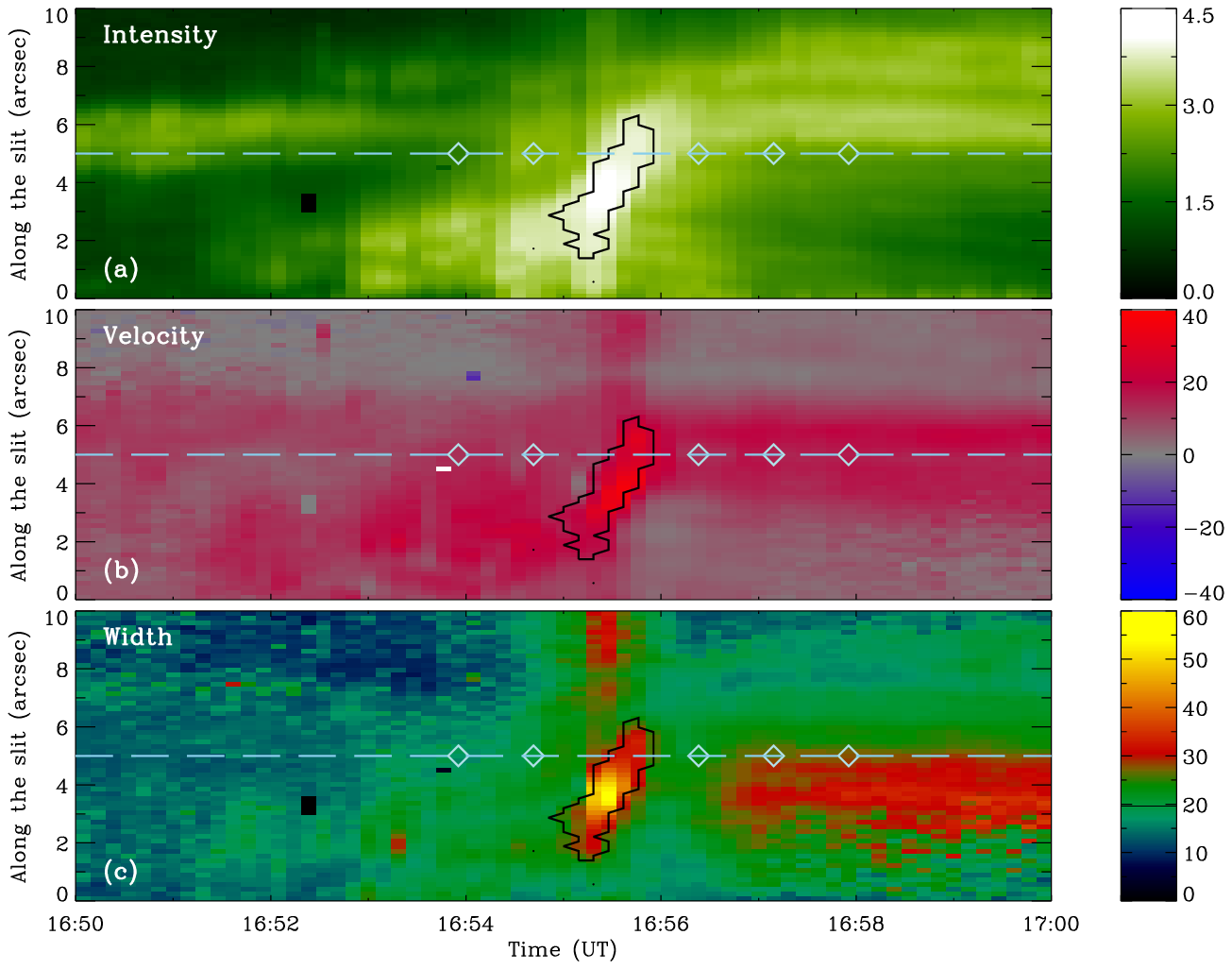}
\caption{{\small Space-time plots of the line intensity (in logarithmic units of DNs), Doppler velocity (in units of km s$^{-1}$), and line width (in units of km s$^{-1}$) of the \siiv~line derived from the moment analysis for flare 1. The horizontal dashed line denotes the location at which the line profiles are analyzed, and the diamonds represent the sample times selected. The contour in each panel delineates the region of saturation.  }}
\label{fig-mo1}
\end{figure*}

Figure \ref{fig-mo1} shows the space-time diagrams of the total intensity, Doppler velocity, and line width of \siiv~derived from the moment method. Note that for the saturation region (marked by a black contour), we still provide the results of  the moment analysis for reference \footnote{In fact, we did some tests by truncating the unsaturated line profiles to mimic the saturated cases. It is found that applying the moment analysis to the saturated \siiv~line profiles at flare ribbons seems to be OK especially for just tracing the evolving trend of the physical parameters.} (similarly for flare 2 as described in Section \ref{fou-res2}). As seen from Figure \ref{fig-mo1}(a), the \siiv~intensity increases gradually with the occurrence of the flare, and the flare ribbon undergoes an apparent motion toward northwest from $\sim$16:53 UT. Mostly, the \siiv~line shows notable redshifts as revealed from the velocity map. The redshift velocity can reach about 40 km s$^{-1}$, which is within the typical velocity range of chromospheric condensation. It is also found that the redshift velocity keeps over tens of km s$^{-1}$ for some minutes after the intensity peak time. In addition, the \siiv~line width shows a similar trend to the Doppler velocity, which is also a typical feature for the chromospheric condensation plasma \citep{liyd15} \\

\begin{figure*}
\centering
\includegraphics[width=19cm, height=10cm]{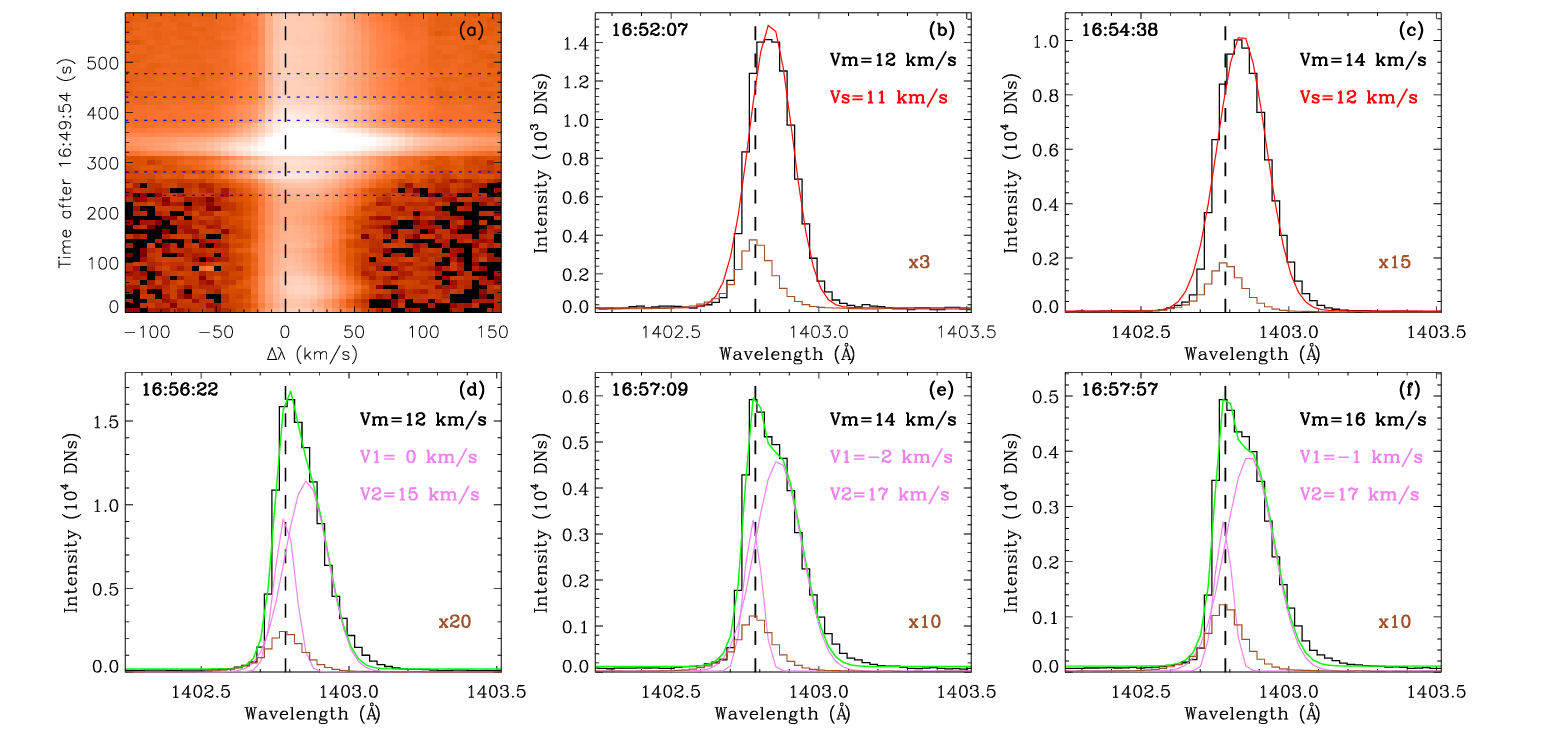}
\caption{{\small (a) Temporal evolution of the \siiv~1402.77 {\AA} line at a ribbon location (marked by the horizontal line in Figure \ref{fig-mo1}). The vertical line indicates the reference wavelength. (b)--(f) Line profiles of the \siiv~1402.77 {\AA} line at 5 times corresponding to the 5 horizontal lines in (a) and also the 5 diamonds in Figure \ref{fig-mo1}. Also plotted on (b)--(f) are the results of either single-Gaussian fitting or double-Gaussian fitting with the line styles the same as in Figure \ref{fig-met}. The line profile plotted in brown refers to the reference profile averaged over a relatively quiet region, multiplied by a factor shown in the lower right corner.}}
\label{fig-sp1}
\end{figure*}

In Figure \ref{fig-sp1} we plot the temporal evolution of \siiv~spectra as well as some example profiles at a selected ribbon location marked in Figure \ref{fig-mo1}. It is seen that, before the \siiv~intensity peak time, the profiles are symmetric and show a good Gaussian shape; then we apply a single Gaussian function to fit them (see Figures \ref{fig-sp1}(b) and (c)). However, after the \siiv~intensity peak, the profiles appear to be asymmetric, or the red wing is enhanced compared with the blue wing (Figures \ref{fig-sp1}(d)--(f)); then we make a double Gaussian fitting for such profiles, which are decomposed into a relatively static component and a redshifted one. The fitting results show that the redshifted component is stronger and broader than the static one. \\ 

\begin{figure*}
\centering
\includegraphics[width=16cm, height=12cm]{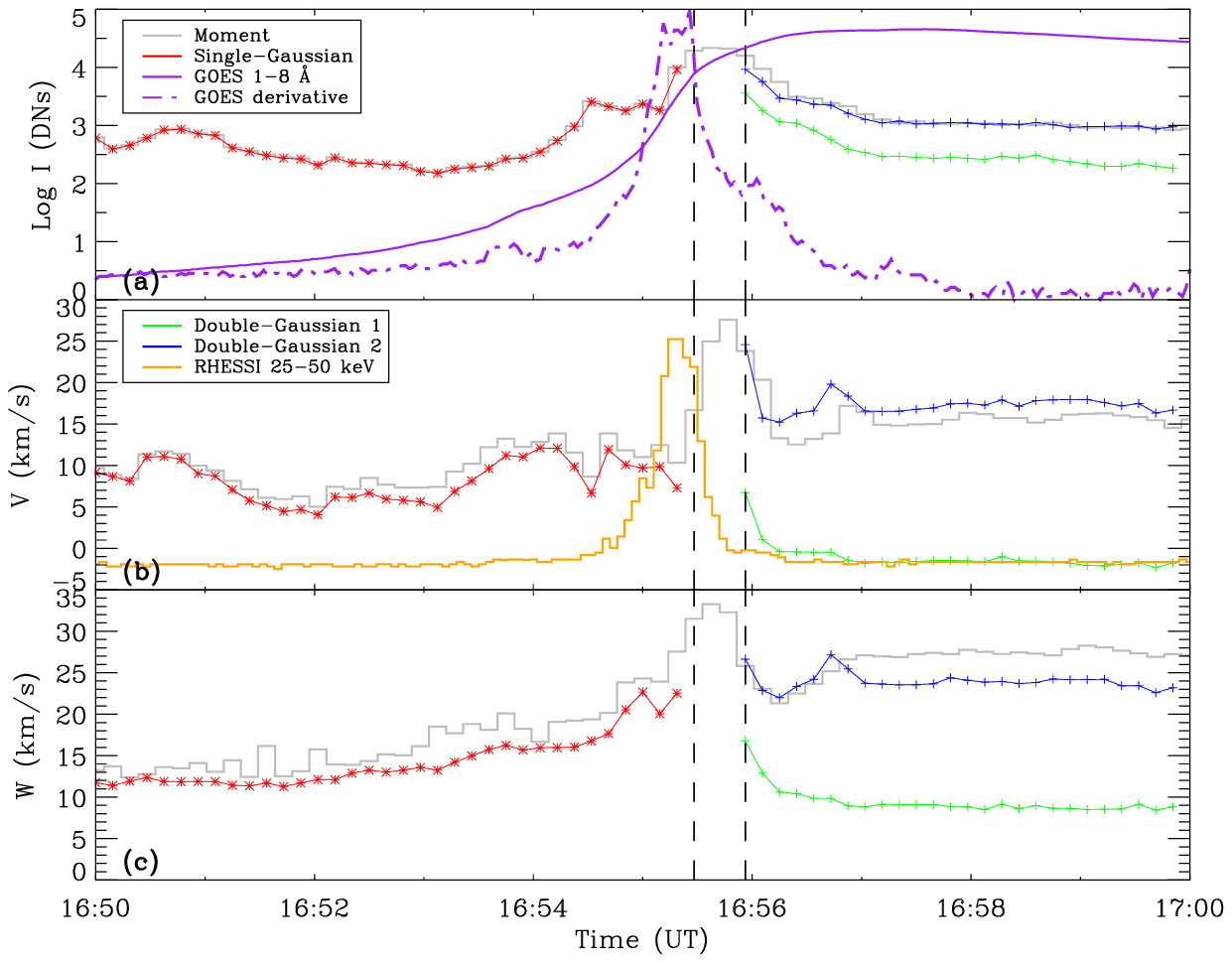}
\caption{{\small Time evolutions of (a) the line intensity, (b) Doppler velocity, and (c) line width of the \siiv~1402.77 {\AA} line at a ribbon location (marked in Figure \ref{fig-mo1}) in flare 1. Also plotted are the {\em GOES} 1--8 {\AA} flux and its time derivative on panel (a) and the {\em RHESSI} HXR emission at 25--50 keV on panel (b). The two vertical dashed lines show the time period of saturation. The gray lines represent the results from moment analysis. }}
\label{fig-f1}
\end{figure*}

Figure \ref{fig-f1} plots the \siiv~line parameters deduced from the single or double Gaussian fitting at the selected ribbon location, together with the {\em GOES} SXR and {\em RHESSI} HXR light curves for comparison. It is clearly seen that, before the \siiv~intensity peak time (also around the HXR peak time), the single Gaussian fitting and the moment method yield almost the same results. During this period, the \siiv~intensity increases with time and the redshift velocity also increases somewhat from 4 to 13 km s$^{-1}$. We notice that the line width increases gradually as well. After that, the line profiles become asymmetric rather than wholly shifted and the Doppler velocity for the redshifted component is about 20 km s$^{-1}$. By comparison, the velocity of the relatively static component is almost zero in this case. The wholly redshifted profile and the subsequent redshifted component are supposed to be caused by a downflow in the line formation layer, which is known as chromospheric condensation. Note that if we still apply the single Gaussian fitting to the asymmetric line profiles, then the downflow velocity would be underestimated, as mentioned in Section \ref{thi-met}. \\

\subsection{Flare 2: the X1.6 flare on 2014 September 10}
\label{fou-res2}
\subsubsection{Observation Overview}

\begin{figure*}
\centering
\includegraphics[width=16cm, height=12cm]{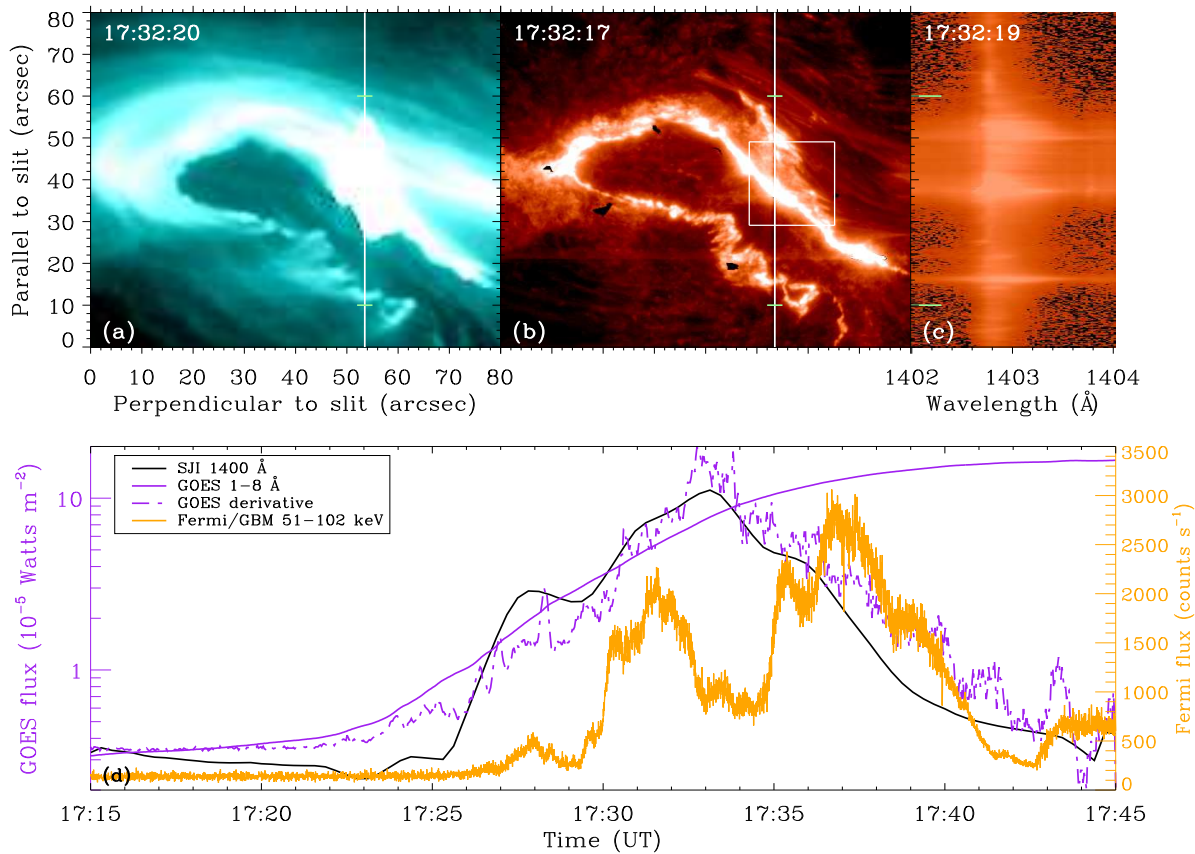}
\caption{{\small (a) {\em SDO}/AIA 131 {\AA} image taken at 17:32:20 UT on 2014 September 10 showing the morphology of flare 2 from a sigmoid region. (b) {\em IRIS}/SJI 1400 {\AA} image taken at 17:32:17 UT. (c) {\em IRIS} slit spectra of the \siiv~1402.77 {\AA} line taken at 17:32:19 UT. In panels (a) and (b), the vertical white line indicates the slit position and the two green bars enclose the part of the slit that is shown in Figure \ref{fig-mo2}. (d) Light curves of the SJI 1400 {\AA} intensity (black, arbitrary scale) averaged over the flare region (marked by the white square), {\em GOES} 1--8 {\AA} flux (purple solid) and its time derivative (purple dash-dotted), and the {\em Fermi} GBM 51--102 keV flux (orange).}}
\label{fig-fer}
\end{figure*}

The X1.6 flare on 2014 September 10 is from a sigmoid region in NOAA AR 12158. The flare started at 17:21 UT and peaked at 17:45 UT. Here we mainly focus on the time period before the flare peak, i.e., the rise phase of the flare. The {{\em IRIS}} SJIs were taken at a cadence of $\sim$19 s at 1400 and 2796 {\AA} and the exposure time changed from $\sim$8 s to 2.4 s after $\sim$17:27 UT. Since the {\em RHESSI} data are not available for this flare, we use the observation of {\em Fermi} GBM instead. Figures \ref{fig-fer}(a)--(c) show the AIA 131 {\AA} image, the SJI at 1400 {\AA}, and the slit spectra of \siiv, respectively. We can see that the {\em IRIS} slit crossed the eastern flare ribbon at two locations. Note that this flare is so energetic that many pixels are saturated. The {\em GOES} SXR light curve and its time derivative, as well as the {\em Fermi} GBM 51--102 keV and SJI 1400 {\AA} light curves, are shown in Figure \ref{fig-fer}(d). It is seen that the Neupert effect seems also valid for this flare. In particular, when the HXR flux keeps growing, the SJI 1400 {\AA} emission rises and then reaches its maximum.\\

\subsubsection{Results}

\begin{figure*}
\centering
\includegraphics[width=16cm, height=12cm]{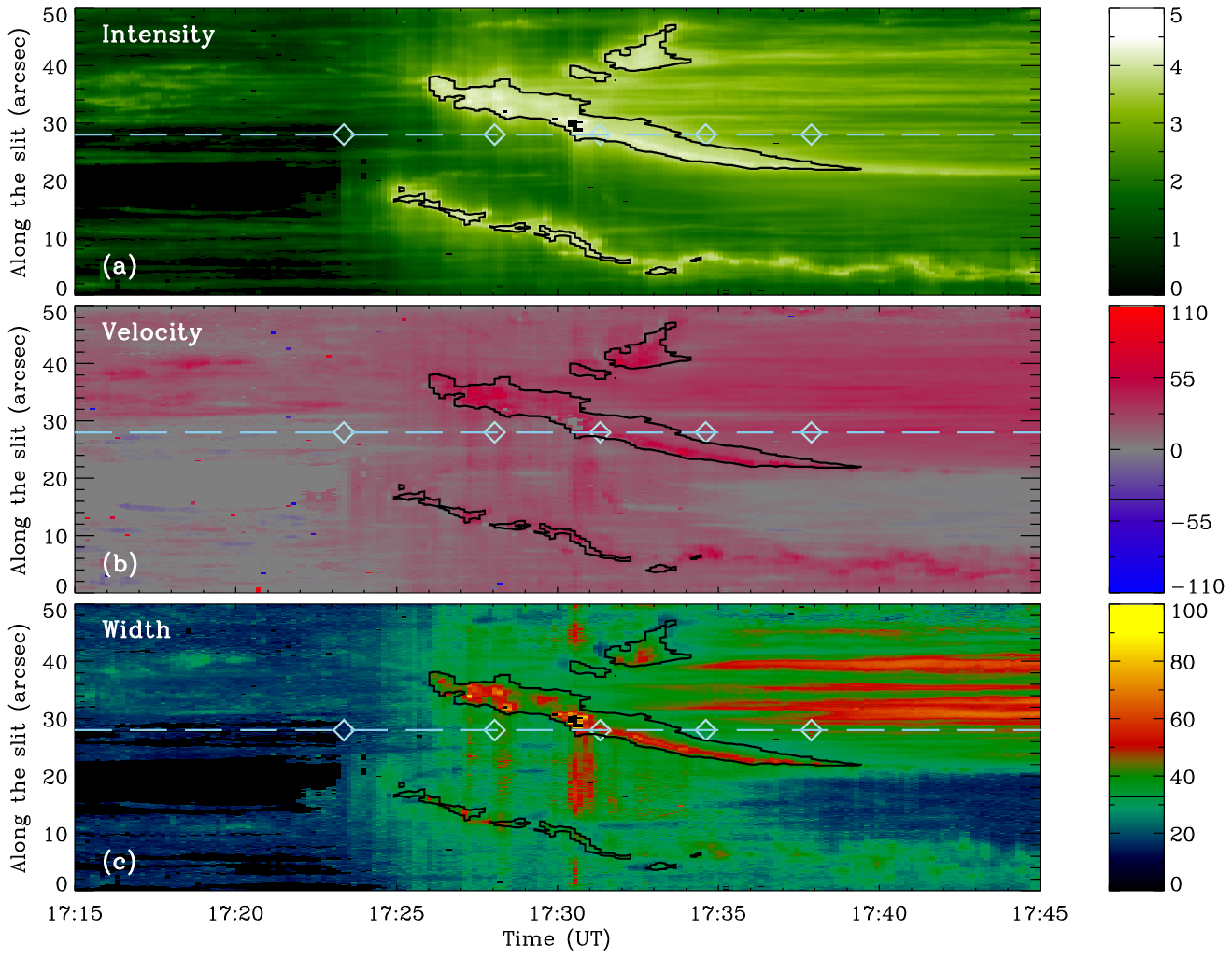}
\caption{{\small Space-time plots of the line intensity (in logarithmic units of DNs), Doppler velocity (in units of km s$^{-1}$), and line width (in units of km s$^{-1}$) of the \siiv~line derived from the moment analysis for flare 2. The horizontal dashed line denotes the location at which the line profiles are analyzed, and the diamonds represent the sample times selected. The contour in each panel delineates the region of saturation. }}
\label{fig-mo2}
\end{figure*}

The space-time diagrams of the total intensity, Doppler velocity, and line width of \siiv~for flare 2 are shown in Figure \ref{fig-mo2}. The intensity map reveals an apparent motion toward south of two brightening features that correspond to the two locations at the flare ribbon crossed by the {\em IRIS} slit. In the following study, we focus on the upper (north) ribbon location. It is seen that when the flare begins, the intensity, velocity, and width of the line increase rapidly at the same time. The region that we are interested in displays significant redshifts in the \siiv~line. \\ 

\begin{figure*}
\centering
\includegraphics[width=19cm,height=10cm]{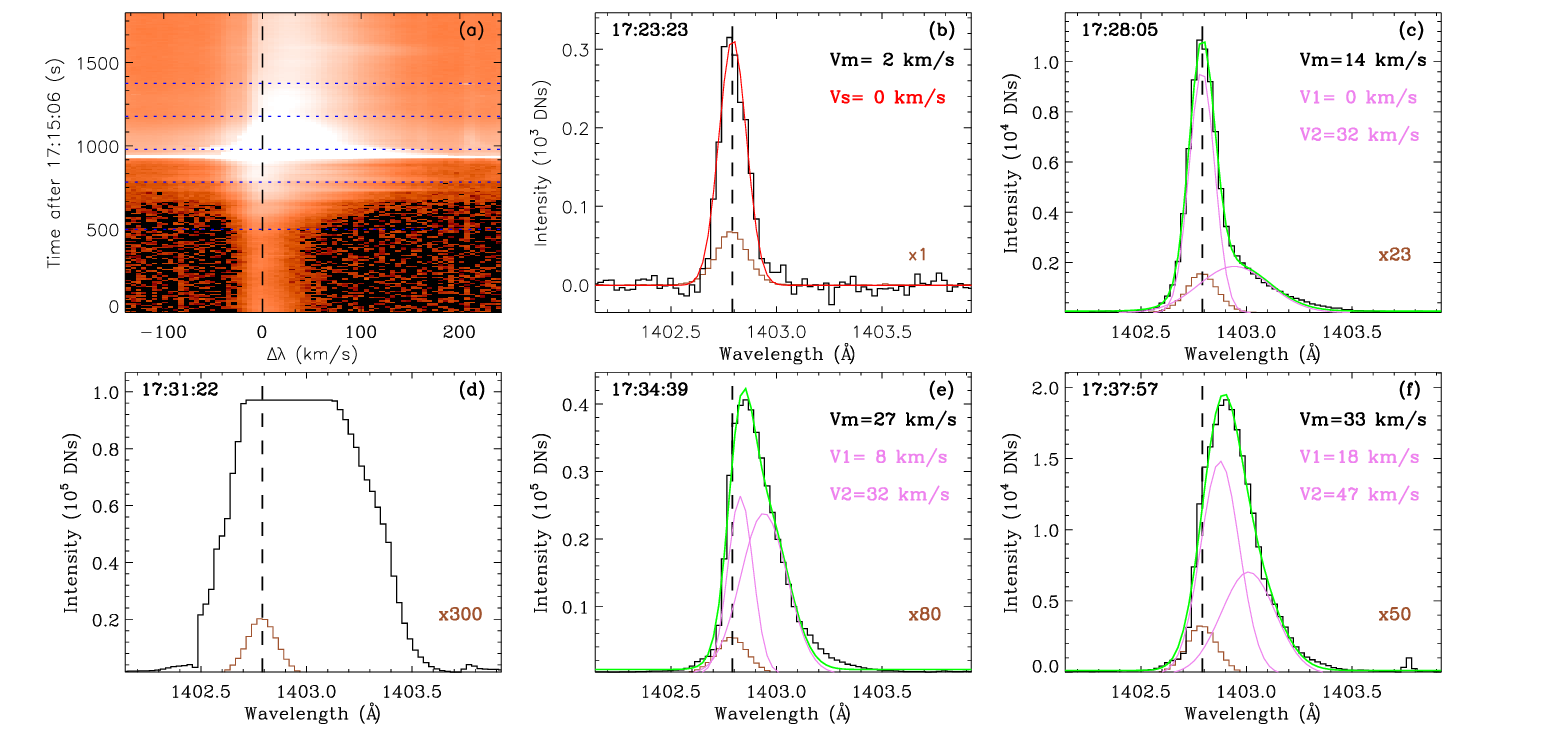}
\caption{{\small (a) Temporal evolution of the \siiv~1402.77 {\AA} line at a ribbon location (marked by the horizontal line in Figure \ref{fig-mo2}). The vertical line indicates the reference wavelength. (b)--(f) Line profiles of the \siiv~1402.77 {\AA} line at 5 times corresponding to the 5 horizontal lines in (a) and also the 5 diamonds in Figure \ref{fig-mo2}. Also plotted on (b)--(f) are the results of either single-Gaussian fitting or double-Gaussian fitting with the line styles the same as in Figure \ref{fig-met}. The line profile plotted in brown refers to the reference profile averaged over a relatively quiet region, multiplied by a factor shown in the lower right corner. Note that panel (d) shows an example of saturation that we neglect when analyzing the line profiles.}}
\label{fig-sp2}
\end{figure*}

In this flare, an evident line asymmetry appears before the \siiv~intensity peak time. Figure \ref{fig-sp2} plots some typical \siiv~line profiles at a selected ribbon location. It is seen that a red-wing enhancement appears very soon after the flare begins. In spite of the saturation effect, one can still notice that the red-wing enhancement increases with time, as revealed from the double Gaussian fitting. After the \siiv~intensity peak, the strength of the redshifted component decreases but the velocity still increases for a while. The line profiles gradually become to be wholly redshifted with a small asymmetry during this period. \\

\begin{figure*}
\centering
\includegraphics[width=16cm, height=12cm]{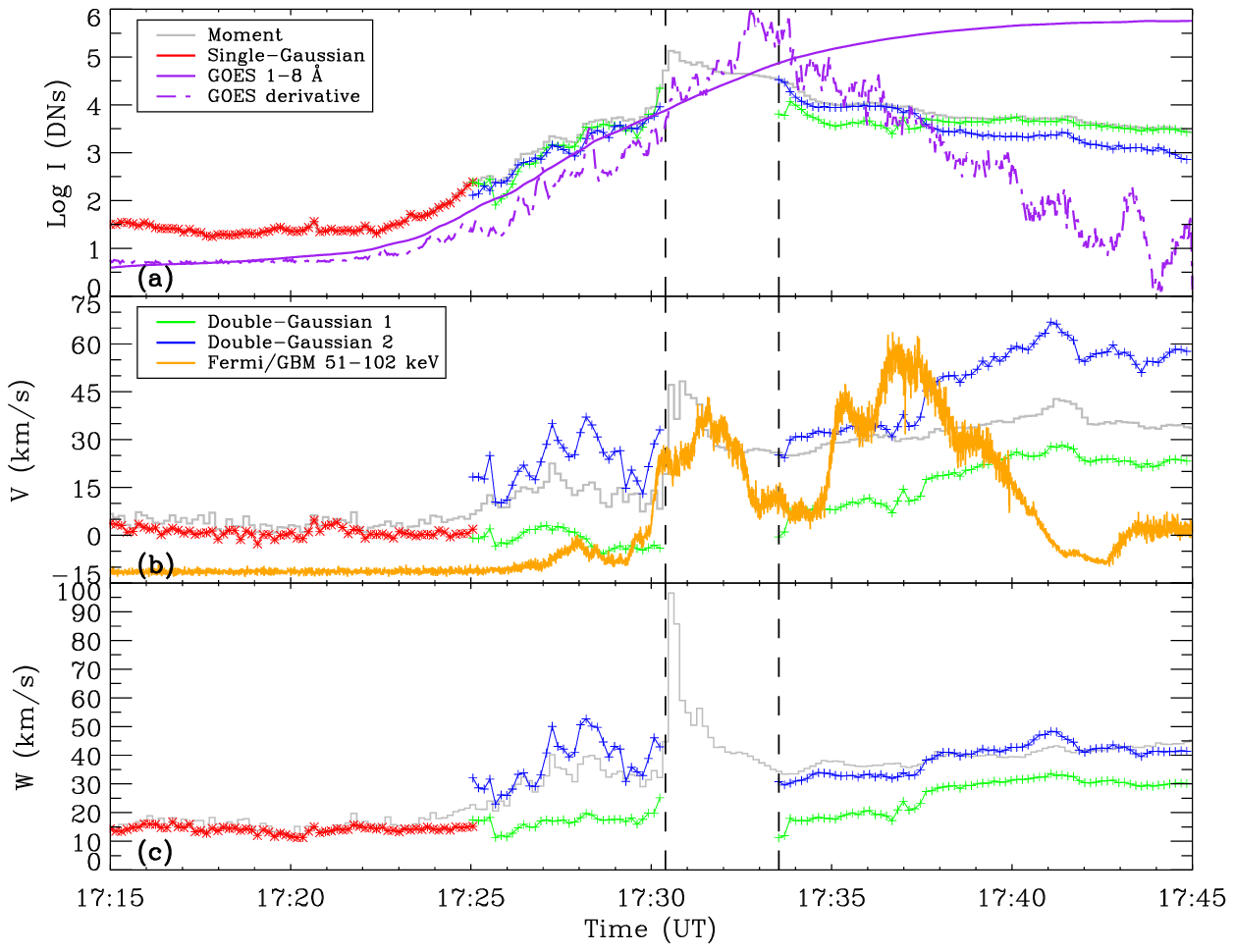}
\caption{{\small Time evolutions of the (a) line intensity, (b) Doppler velocity, and (c) line width of the \siiv~1402.77 {\AA} line at a ribbon location (marked in Figure \ref{fig-mo2}) in flare 2. Also plotted are the {\em GOES} 1--8 {\AA} flux and its time derivative on panel (a), and the {\em Fermi} GBM 51--102 keV flux on panel (b). The two vertical dashed lines show the time period of saturation. The gray lines represent the results from moment analysis.}}
\label{fig-f2}
\end{figure*}

Figure \ref{fig-f2} shows the temporal evolution of the \siiv\ line parameters obtained from either the single or double Gaussian fitting at the selected ribbon location, along with the {\em GOES} SXR and {\em Fermi} GBM 51--102 keV light curves. It is seen that the \siiv~intensity has a rise-and-fall evolution during the rise of the SXR emission, while the Doppler velocity and the line width keep increasing with time. It is worth noting that the velocity of the relatively static component reaches almost 30 km s$^{-1}$ at a late time ($\sim$17:41 UT). The Doppler velocity of the redshifted component can be up to about 70 km s$^{-1}$. We also note that the \siiv~redshifts appear around the same time as the rising of the HXR emission. \\ 

\subsection{Flare 3: the B1.8 flare on 2016 December 2}
\label{fou-res3}
\subsubsection{Observation Overview}

\begin{figure*}
\centering
\includegraphics[width=16cm, height=12cm]{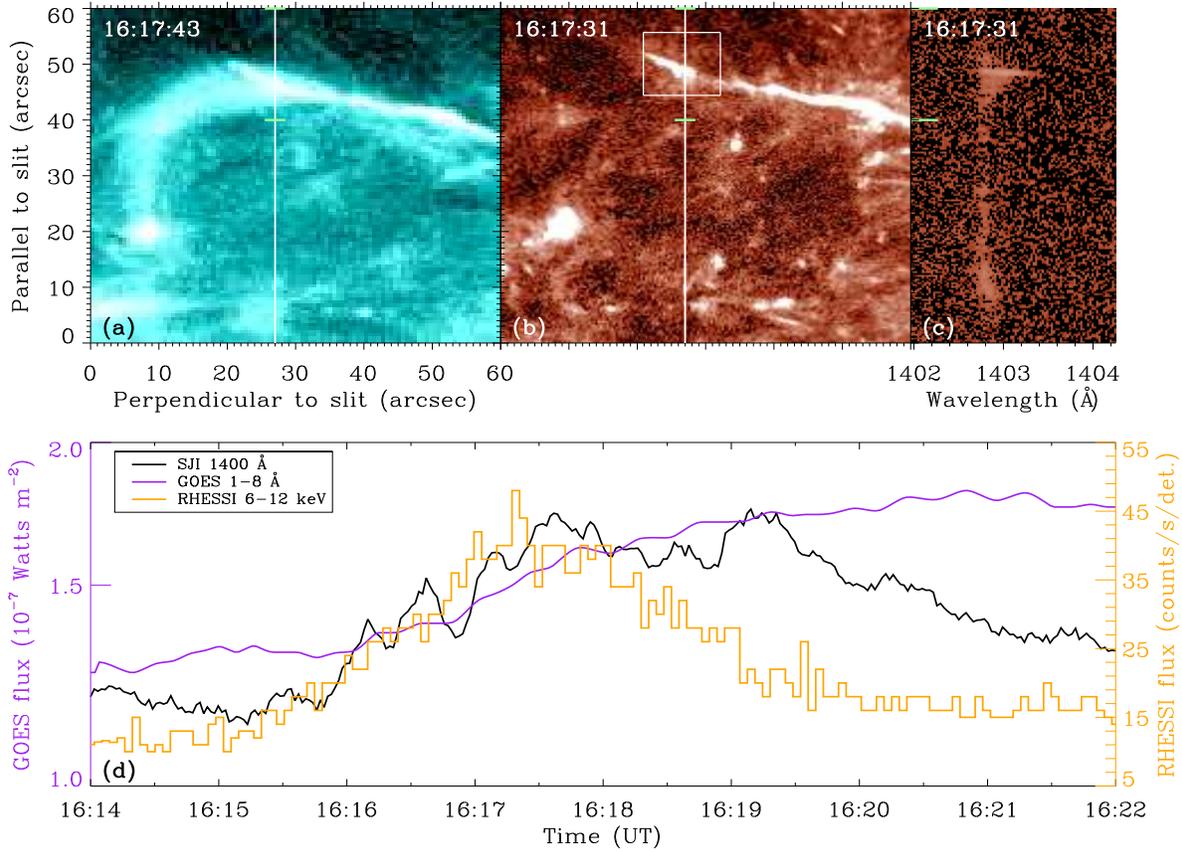}
\caption{{\small (a) {\em SDO}/AIA 131 {\AA} image taken at 16:17:43 UT on 2016 December 2 showing the morphology of flare 3. (b) {\em IRIS}/SJI 1400 {\AA} image taken at 16:17:31 UT. (c) {\em IRIS} slit spectra of the \siiv~1402.77 {\AA} line taken at 16:17:31 UT. Also plotted on panels (a) and (b) are the slit position (vertical white line). The two green bars enclose the part of the slit that is shown in Figure \ref{fig-mo3}. (d) Light curves of the SJI 1400 {\AA} intensity (black, arbitrary scale) averaged over the flare region (marked by the white square), {\em GOES} 1--8 {\AA} flux (purple), and {\em RHESSI} 6--12 keV flux (orange).  }}
\label{fig-rh3}
\end{figure*}

The B1.8 microflare on 2016 December 2 started at $\sim$16:14 UT and peaked at $\sim$16:24 UT. \IRIS\ observed this flare from 16:05 UT to 16:22 UT with a very high cadence of 1.7 s. The SJIs were only taken at 1400 {\AA} with a cadence of 2 s. Figures \ref{fig-rh3}(a)--(c) show the AIA 131 {\AA} image, the SJI at 1400 {\AA}, and the slit spectra of \siiv. One can see that the {\em IRIS} slit crossed the narrow flare ribbon where evident redshifts appear in the \siiv\ spectra. We show the {\em GOES} SXR, {\em RHESSI} 6--12 keV, and SJI 1400 {\AA} light curves in Figure \ref{fig-rh3}(d). It is seen that the SJI 1400 \AA\ emission starts to rise at about the same time as the {\em RHESSI} 6--12 keV emission, both of which reach their maxima before the {\em GOES} SXR peak time. Note that the intensity of the \siiv~line in this microflare is much weaker compared with the former two large flares; therefore, we only choose a part of the flare region with an intensity above a certain threshold for analysis here. \\

\subsubsection{Results}

\begin{figure*}
\centering
\includegraphics[width=16cm, height=12cm]{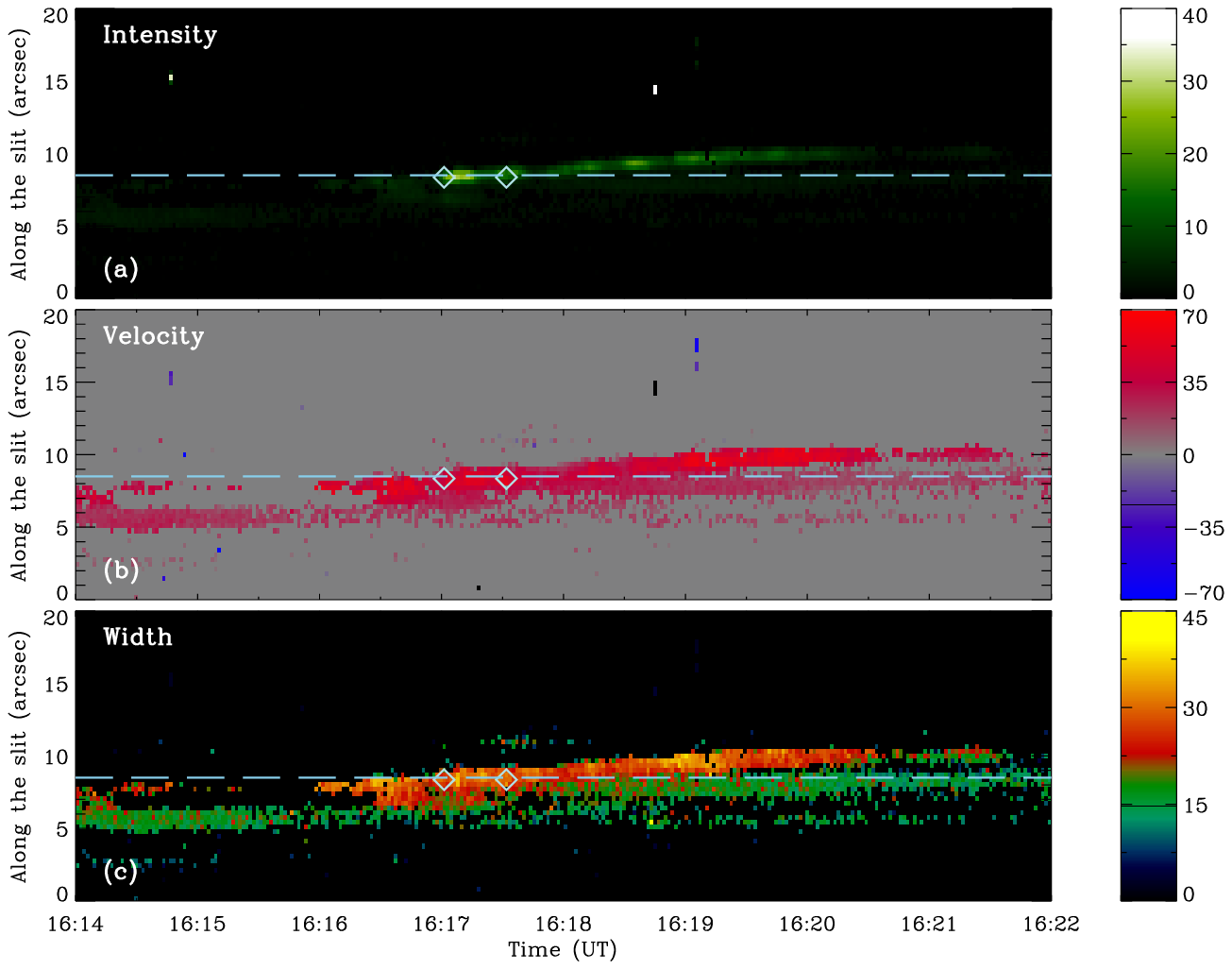}
\caption{{\small Space-time plots of the line intensity (in units of DNs), Doppler velocity (in units of km s$^{-1}$), and line width (in units of km s$^{-1}$) of the \siiv~line derived from the moment analysis for flare 3. The horizontal dashed line denotes the location at which the line profiles are analyzed, and the diamonds represent the sample times selected.  }}
\label{fig-mo3}
\end{figure*}

Figure \ref{fig-mo3} shows the space-time diagrams of the intensity, velocity, and width of the \siiv~line derived from the moment analysis for the region brighter than the intensity threshold. It is seen that, although the increase of the intensity is small, there still shows up an apparent motion of the flare ribbon with time. Note that the redshift of the \siiv~line just appears in a very narrow region of less than 5$^{''}$ along the slit. However, the Doppler velocity can reach up to 60 km s$^{-1}$. In addition, the line width changes consistently with the Doppler velocity as in the former two flares. \\

\begin{figure*}
\centering
\includegraphics[width=19cm,height=7cm]{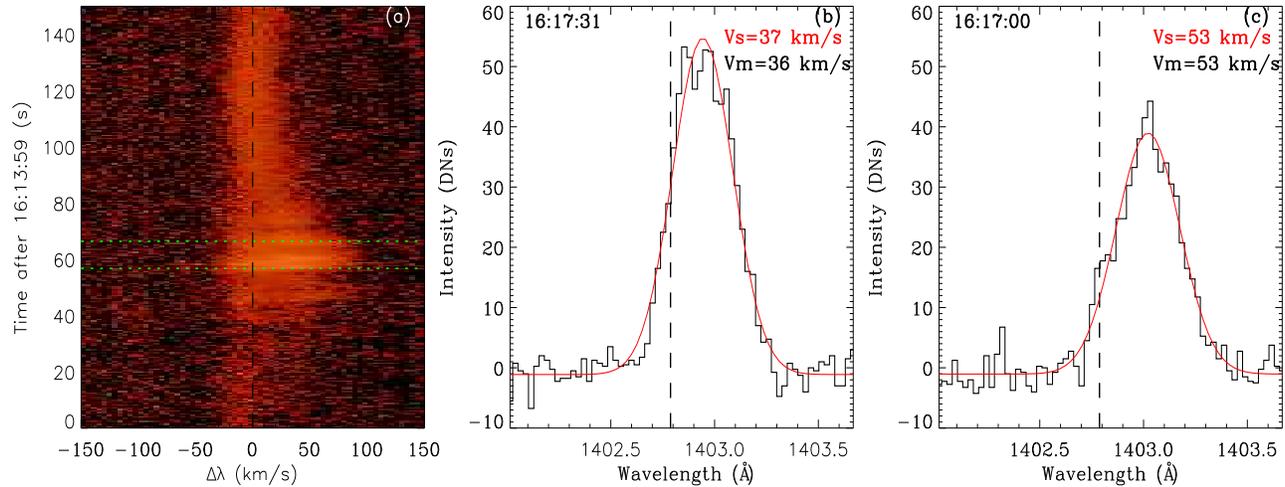}
\caption{{\small (a) Temporal evolution of the \siiv~1402.77 {\AA} line at a ribbon location (marked by the horizontal line in Figure \ref{fig-mo3}). (b)--(c) Line profiles of the \siiv~1402.77 {\AA} line at 2 times corresponding to the 2 horizontal lines in (a) and also the 2 diamonds in Figure \ref{fig-mo3}. Also plotted on panels (b) and (c) are the results of single-Gaussian fitting. The line styles are the same as in Figure \ref{fig-met}.}}
\label{fig-sp3}
\end{figure*}

Compared with the former two large flares, the \siiv~line profiles in this microflare are quite simple, which only display a single Gaussian shape throughout the flare evolution. Figure \ref{fig-sp3} plots two example line profiles at the selected ribbon location. It is seen that the profiles are wholly redshifted during the flare. Thus, the redshift velocity derived from the single Gaussian fitting is almost the same as that obtained by the moment analysis. \\

\begin{figure*}
\centering
\includegraphics[width=16cm, height=12cm]{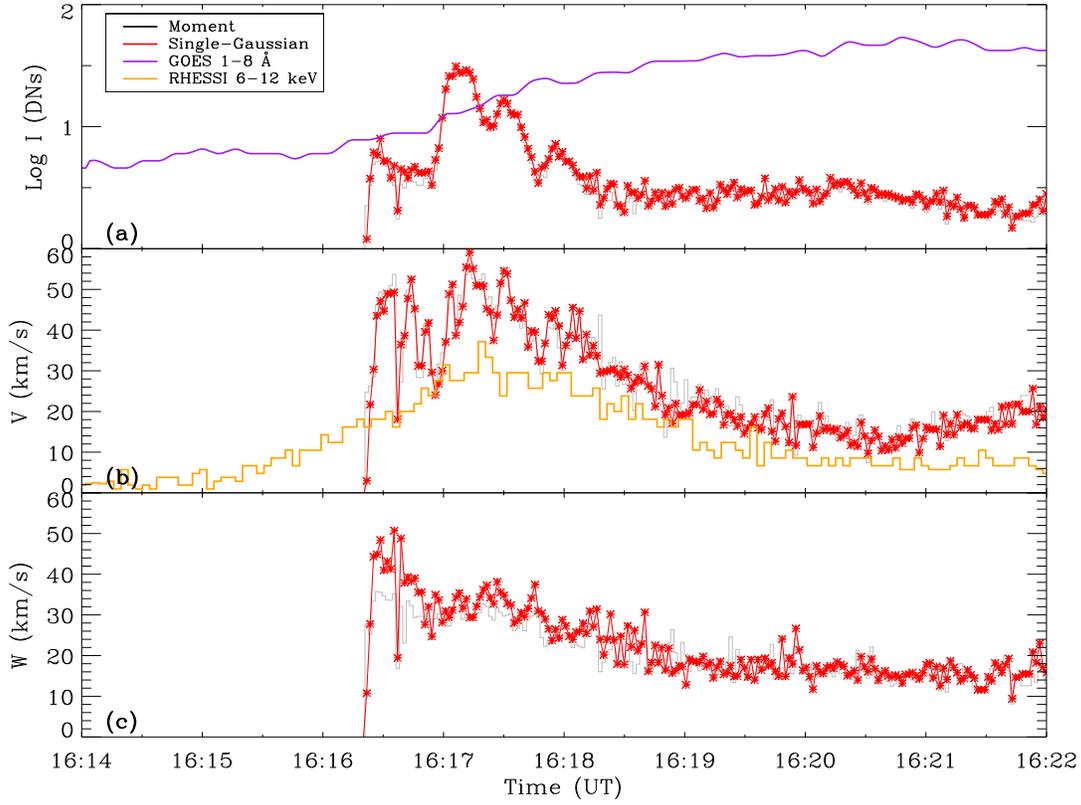}
\caption{{\small Time evolutions of the (a) line intensity, (b) Doppler velocity, and (c) line width of the \siiv~1402.77 {\AA} line at a ribbon location (marked in Figure \ref{fig-mo3}) in flare 3. Also plotted are the {\em GOES} 1--8 {\AA} flux on panel (a) and the {\em RHESSI} 6--12 keV flux on panel (b). The gray lines represent the results from moment analysis.}}
\label{fig-f3}
\end{figure*}

Figure \ref{fig-f3} plots the temporal evolutions of the line parameters derived from the moment analysis and single Gaussian fitting, as well as the {\em GOES} SXR and the {\em RHESSI} 6--12 keV light curves. One can see that the intensity, velocity, and width of the \siiv line rise and decline almost in phase, also in rough coincidence with the {\em RHESSI} 6--12 keV flux. \\

\section{Discussions}
\label{fiv-dis}
There have been published a number of studies on flares 1 and 2 from different aspects \citep{grah15, kuri15, lidn15, tian15}. For example, \citet{tian15} studied the chromospheric evaporation in these two flares by mainly focusing on the \fexxi\ line that is shown to be completely blushifted at the flare ribbons. In particular, the blueshift velocity of the \fexxi~line shows a good temporal relationship to the HXR 25--50 keV emission, suggesting a chromospheric evaporation caused by nonthermal electron heating. \citet{tian15} also reported enhanced emission in the red wings of the \siiv\ line profiles. The Doppler velocity of the redshifts is found to be larger than the previously reported values and lasts for a relatively long time in the decay phase. They ascribed the redshifts in the cool lines to chromospheric condensation and also downward cooling plasma from the corona. In this work, we mainly focus on the rise phase of the flare. The redshifts detected in the \siiv\ line at the flare ribbons are supposed to be a signature of chromospheric condensation. In addition, as pointed out by \citet{liyk17}, some of the \siiv~line profiles at the flare ribbons can be well fitted by a Gaussian shape, but others cannot. Thus, we have distinguished two types of \siiv~line profiles here and apply different methods to fit them accordingly. We also notice that flare 1 has been studied with both observations and radiative hydrodynamic simulations by \citet{kuri15}. The authors found red and blue asymmetries in the H$\alpha$ line but only a weak red asymmetry in the Ca {\sc ii} 8542 {\AA} line at the flare kernel. Resorting to the simulations, they interpreted the blue asymmetry in H$\alpha$ in terms of downflows owing to plasma condensation. This is consistent with our explanations although we detect red asymmetry in the \siiv\ line. Note that the H$\alpha$ line is optically thick, which can show different spectral features from optically thin lines. \\ 

From the HXR emission and its relationship to the \siiv\ line profiles, we could explore the heating mechanisms for the three flares under study. We find that in flares 1 and 2, the HXR emissions can be visible up to 50 keV or 100 keV, respectively, and they roughly show a relationship to the appearance of the asymmetry of the \siiv~line. In flare 3, however, there is no detectable HXR emission above 12 keV or no obvious asymmetry in the \siiv\ line profile. Therefore, we consider that nonthermal electron beam plays a major role in heating the chromosphere in flares 1 and 2, while thermal conduction may be the dominant heating way in flare 3. Both of the heating mechanisms are thought to be reflected in the different behaviors of the \siiv\ line profiles. \\

The fact that the shape of the \siiv\ line profile at the flare ribbon depends on the heating mechanisms has been investigated via numerical simulations. \citet{poli18} performed nanoflare simulations with nonthermal electron beam heating and in situ thermal heating. They found that in the case of thermal heating, the \siiv~line more likely tends to be wholly redshifted, just like flare 3 in our study. In addition, \citet{kerr19} studied two resonance \siiv~lines in flare models with electron beam heating. They reported that the energy flux, the low-energy cutoff, and the spectral index of the nonthermal electron beam could all influence the \siiv~line profiles. For instance, with a higher injected energy flux, the \siiv~line could more likely show a red asymmetry. Such a case seems to be consistent with our observations in flares 1 and 2 that have a relatively high magnitude of HXR flux. Furthermore, with the condensation propagating downward, there possibly exist two regions, a condensation front and a relatively stationary region, both of which contribute to the line emission. This might explain the two components of the line profile with different Doppler velocities. We notice that in the simulations of chromospheric condensation driven by electron beams, \citet{kowa18} revealed two flaring layers in the chromosphere: a chromospheric condensation region with downflowing, hot, and dense plasma as well as a stationary layer below it. Note that the stationary layer can also be heated by the high-energy electrons. The two layers can reproduce the red-wing asymmetry as shown in some chromospheric emission lines during flares. Finally, although the simulations mentioned above could reproduce the \siiv\ spectral features as shown in our work, there still exist some discrepancies between the observations and simulations. For example, blueshifts of the \siiv~line have been displayed in simulations, especially in the cases with a lower injected flux. However, we do not find any blueshifts in the \siiv~line at the flare ribbon in our study. In addition, \citet{kerr19} emphasized the opacity effects in their study. However, our analysis of the \siiv\ line for the three flares is based on the assumption that the line is formed in an optically thin condition. In the future, we will study more flare events as well as carry out radiative hydrodynamic simulations to further investigate the relationship between the \siiv~line profiles and the heating mechanisms.

\section{Conclusions}
\label{six-sum}
By using the high-resolution spectral observations from {\em IRIS}, we have tracked the temporal evolution of the \siiv~1402.77 {\AA} line profiles at the flare ribbons in three flare events. In all of the events, we detect evident redshifts in this line, indicative of downward mass motions in the TR layer. The line intensity, Doppler velocity, and line width increase and reach their peaks at nearly the same time. It is interesting that in the two large flares, the \siiv~profiles can transit from wholly redshifted to red-wing enhanced ones in the rise phase of the flare; however, in the third microflare, the \siiv~profiles are wholly redshifted throughout the flare process. We have performed either a single or double Gaussian fitting to the line profiles. Generally speaking, the single Gaussian fitting is appropriate for dealing with the wholly shifted line profiles while it would underestimate the Doppler velocities in the case of red-asymmetric profiles. In the latter case, the double Gaussian fitting should be adopted, which can reveal two emission components, a relatively static one and a redshifted one for most cases. Specifically, the redshifted component is thought to originate from the region with a significant downward velocity. The downflow velocities corresponding to the redshifted components or the redshifts as a whole are measured to be a few tens of km s$^{-1}$. These values are among the typical speed of chromospheric condensation reported in previous studies. Existence of an observationally prominent chromospheric condensation implies that these flares belong to the events of explosive evaporation. \\

In order to explore the physical mechanism behind the \siiv\ line profiles, we analyze the HXR emission for the three flares. We find that in the two large flares, the HXR emissions are prominent, which can be visible up to 50 keV or even 100 keV. In particular, these HXR emissions match in time with the development of the \siiv\ redshifts and even the red asymmetries to some extent. In the microflare, however, the HXR emission is very weak and only visible up to 12 keV, and the \siiv\ line appears to be wholly redshifted during the flare. Thus, we consider that nonthermal electron beam plays a major role in heating the plasma in the two large flares, while thermal conduction may be the dominant heating mechanism in the microflare. \\


\acknowledgments
The authors would like to thank the anonymous referee for constructive comments. We also thank Dr. Hui Tian for helpful discussions. {\em IRIS} is a NASA small explorer mission developed and operated by LMSAL with mission operations executed at NASA Ames Research Center and major contributions to downlink communications funded by the Norwegian Space Center (NSC, Norway) through an ESA PRODEX contract. {\em SDO} is a mission of NASA's Living With a Star Program. The project is supported by NSFC under grants 11733003, 11873095, 11903020, 11533005, 11961131002, and U1731241. Y.L. is also supported by CAS Pioneer Talents Program for Young Scientists and XDA15052200, XDA15320103 and XDA15320301.

\bibliographystyle{apj}

\end{document}